\RequirePackage{fix-cm}
\documentclass[smallextended]{svjour3}       
\smartqed  
\usepackage{graphicx}

\usepackage{amsmath,amsfonts,amssymb}
\usepackage{graphicx}
\usepackage{setspace}
\usepackage{tocloft}
\usepackage{subfig}
\usepackage{color,xcolor} 
\usepackage{multirow}
\usepackage{float}

\begin{document}

\title{Dense U-net for single image super-resolution with shuffle pooling layer
}


\author{Zhengyang Lu        \and
        Ying Chen 
}


\institute{Zhengyang Lu \at
              Jiangnan University, Key Laboratory of Advanced Process Control for Light Industry (Ministry of Education),  Wuxi, China
              \email{7191905018@stu.jiangnan.edu.cn}           
           \and
           Ying Chen \at
              Jiangnan University, Key Laboratory of Advanced Process Control for Light Industry (Ministry of Education),  Wuxi, China
	\email{chenying@jiangnan.edu.cn}     
}

\date{Received: date / Accepted: date}

\maketitle

\begin{abstract}
Recent researches have achieved great progress on single image super-resolution(SISR) due to the development of deep learning in the field of computer vision.
In these method, the high resolution input image is down-scaled to low resolution space using a single filter, commonly max-pooling, before feature extraction.
This means that the feature extraction is performed in biased filtered feature space.
We demonstrate that this is sub-optimal and causes information loss.
In this work, we proposed a state-of-the-art convolutional neural network method called Dense U-net with shuffle pooling. 
To achieve this, a modified U-net with dense blocks, called dense U-net, is proposed for SISR. 
Then, a new pooling strategy called shuffle pooling is designed, which is aimed to replace the dense U-Net for down-scale operation. 
By doing so, we effectively replace the handcrafted filter in the SISR pipeline with more lossy down-sampling filters specifically trained for each feature map, whilst also reducing the information loss of the overall SISR operation.
In addition, a mix loss function, which combined with Mean Square Error(MSE), Structural Similarity Index(SSIM) and Mean Gradient Error (MGE), comes up to reduce the perception loss and high-level information loss.
Our proposed method achieves superior accuracy over previous state-of-the-art on the three benchmark datasets: SET14, BSD300, ICDAR2003. Code is available online\footnote{This super-resolution project is coded by PyTorch and is on {https://www.github.com/MnisterLu/DenseSR}}.
\end{abstract}

\section{Introduction}
\label{sect:intro}  
Single image super-resolution can be used as an effective data argumentation method for most image processing tasks, such as object detection, object recognition, target tracking and instance segmentation. However, various image types under different complex environments impose great challenges for super-resolution in real-world applications.

The global SR problem assumes high-resolution feature map has glut redundant information which treated as noisy and culled in down-sampling operation. It is a highly necessary to be concerned with this information loss problem due to directly remove pixels in down-sampling such as max-pooling and average-pooling.

Many methods assume that the deeper convolutional network have better ability to extract semantic and perceptual features. 
However, not only the first proposed super-resolution method Super-Resolution Convolution Neural Network ({\bf SRCNN})~\cite{dong2015image}, but also the following representative methods, which includes Very Deep network for Super-Resolution ({\bf VDSR})~\cite{kim2016accurate}, Deeply-Recursive Convolutional Networks ({\bf DRCN})~\cite{kim2016deeply} and Deep Back-Projection Networks ({\bf DBPN})~\cite{haris2018deep}, is committed to designing deeper networks to obtain more perceptual features with a single-channel feature extraction and single-channel up-sampling module.
By doing so, these methods largely ignored the loss of low-level texture information since the single-channel overemphasizes high-level semantic information.

In short, previous works inherently limited to defects in the information transmission structure. 
Take U-net as an example, first, the skip-connections between the same depth convolution layers are only one-way. 
Second, traditional max-pooling method caused more than ($\frac{2^{k}-1}{2^{k}}$) of information loss in each down-sampling block when down-scale is $k$ due to directly drop ($2^{k}-1$) pixel values. 
Third, previous methods focus on evaluation method of Peak Signal to Noise Ratio (PSNR), which result in the high performance on PSNR, but largely ignored the gradient error.
In general, previous methods lacked lossless mechanisms to handle information transfer processing.
To solve these problems, we propose the {\bf Dense UnetSR}, which combines the Dense U-net for super-resolution with shuffle pooling method, that greatly enhances the capability of the information transfer mechanism. 

To solve these problems, this paper has the following four contributions:

\begin{enumerate}
\item[1)] we propose a state-of-the-art convolutional neural network method called Dense U-net with shuffle pooling. 
Compare with U-net~\cite{ronneberger2015u} for SISR, Dense U-net enhances the skip-connection part by transmitting all feature layers from different depth to each up-sampling layer. It reduces the information transmission loss because the up-sampling block combines all down-sampling feature maps from all depth of contracting blocks.

\item[2)] A novel pooling method called shuffle pooling is designed for the Dense U-net, which can effectively replace the handcrafted filter in the SISR pipeline with more lossy down-sampling filters specifically trained for each feature map, whilst also reducing the information loss of the overall SISR operation.
As shown in Fig.~\ref{fig:diffarrange}, shuffle pooling module consists of 2 steps: 1) Sampling the input feature map by specified intervals; 2) Reshaping sampled values into down-sampled feature maps.

\item[3)] The mix loss function, which combined with Mean Square Error(MSE)~\cite{huynh2008scope}, Structural Similarity Index (SSIM)~\cite{hore2010image} and Mean Gradient Error(MGE)~\cite{lu2019single}, basically solves the perception loss and high-frequency information loss. Besides MSE loss, the MixE loss considers SSIM loss which helps to better restore brightness, contrast and structure, and MGE loss which helps to better restore sharpness of images. 

\item[4)] The proposed {\bf Dense UnetSR} outperforms the state-of-the-art the SISR methods in SET14, BSD300, ICDAR2003 datasets, especially in the text tasks.

\end{enumerate}

\section{Related work}

He {\textit{et al.}}~\cite{dong2015image} introduced the first deep learning method for single image super-resolution (SISR) task called Super-Resolution Convolution Neural Network ({\bf SRCNN}).  {\bf SRCNN} simply constructs a multi-layer convolutional network directly without considering transmission loss and computational complexity.

In order to reduce the computational complexity of the {\bf SRCNN}, Fast Super-Resolution Convolutional Neural Network ({\bf FSRCNN})~\cite{dong2016accelerating} built a lightweight framework to solve the real-time problem.  {\bf FSRCNN} 
Instead of taking interpolated image by Bicubic\cite{de1962bicubic} as input, {\bf FSRCNN} directly put the LR image into the network, and up-scaled by deconvolution layer finally. Although {\bf FSRCNN}  has reduced the running time of the {\bf SRCNN} by more than 70\%, its low accuracy made it difficult to be applied in real scenes.

Efficient Sub-Pixel Convolutional Neural ({\bf ESPCN})~\cite{shi2016real} introduced an efficient sub-pixel convolution layer which learns an array of up-scaling filters to upscale the low-resolution image into the high-resolution output.
Inspired by ResNet~\cite{he2016deep}, Very Deep network for Super-Resolution ({\bf VDSR})~\cite{kim2016accurate} was proposed to solve the training problem of deeper super-resolution networks.

{\bf VDSR} built an end-to-end network with 20 convolution layers by cascading small filters many times in a deep network structure. 
Super-Resolution Generative Adversarial Networks ({\bf SRGAN})~\cite{ledig2017photo} were the first Generative Adversarial Networks ({\bf GAN})~\cite{goodfellow2014generative} for super-resolution task the first to consider the human subjective evaluation of reconstructed images.  In order to improve the accuracy and perceptual evaluation, {\bf SRGAN} was extremely hard to converge during the training process.

Enhanced Deep Residual Networks for Super-Resolution ({\bf EDSR})~\cite{lim2017enhanced} achieved the highest score in the NTIRE2017 Super-Resolution Challenge Competition~\cite{Agustsson2017NTIRE}. The most significant performance improvement of {\bf EDSR} was to remove the redundant modules of SRResNet which is the generator network of {\bf SRGAN}~\cite{ledig2017photo}, so that the size of the model can be enlarged to improve the image quality.

Deeply-Recursive Convolutional Networks ({\bf DRCN})~\cite{kim2016deeply} and Deep Back-Projection Networks ({\bf DBPN})~\cite{haris2018deep} exploited iterative up-sampling and down-sampling and {\bf DBPN} had better performance because it provided an error feedback mechanism. 
Although {\bf DRCN} and {\bf DBPN} reduced the number of parameters, these methods did not actually solve the problem of high computational complexity due to too much repetitive calculations.

Image Super-Resolution Using Very Deep Residual Channel Attention Networks({\bf RCAN})\cite{zhang2018image} reached much deeper by residual in residual structure than other CNN-based methods and obtained a better performance. The long and short skip connections in residual in residual structure helped to bypass abundant low-frequency information and made the main network learn more effective information. 
Then, channel attention mechanism is proposed to adaptively rescale features by considering interdependencies among feature channels.

To solve the problem of Hindering the representational power of CNNs caused by neglecting to explore the feature correlations of intermediate layers, the second-order attention network({\bf SAN})\cite{dai2019second} was proposed for for more powerful feature expression and feature correlation learning. 
Specifically, the novel trainable second-order channel attention module was developed to adaptively rescale the channel-wise features by using second-order feature statistics for more discriminative representations. 
Furthermore, Dai\cite{dai2019second} presented a non-locally enhanced residual group structure, which not only incorporates non-local operations to capture long-distance spatial contextual information, but also contains repeated local-source residual attention groups to learn increasingly abstract feature representations.

The U-net for Super-Resolution ({\bf UnetSR})~\cite{lu2019single} modified the basic U-net architecture to adapt to the field of SISR and proposed a mix gradient loss function to enhance the sharpness of reconstructed images. 
Although the problem of image blur is considered, {\bf UnetSR} still has many defects in network transmission.

\section{Approach}
\label{sec:1}
In this section, the architecture of Dense U-net for super-resolution with shuffle pooling layer will be described as three parts: 1) Dense U-net network; 2) Shuffle pooling method; 3) Mix loss function.

\begin{figure*}[!h]
\centerline{\includegraphics[width=\columnwidth]{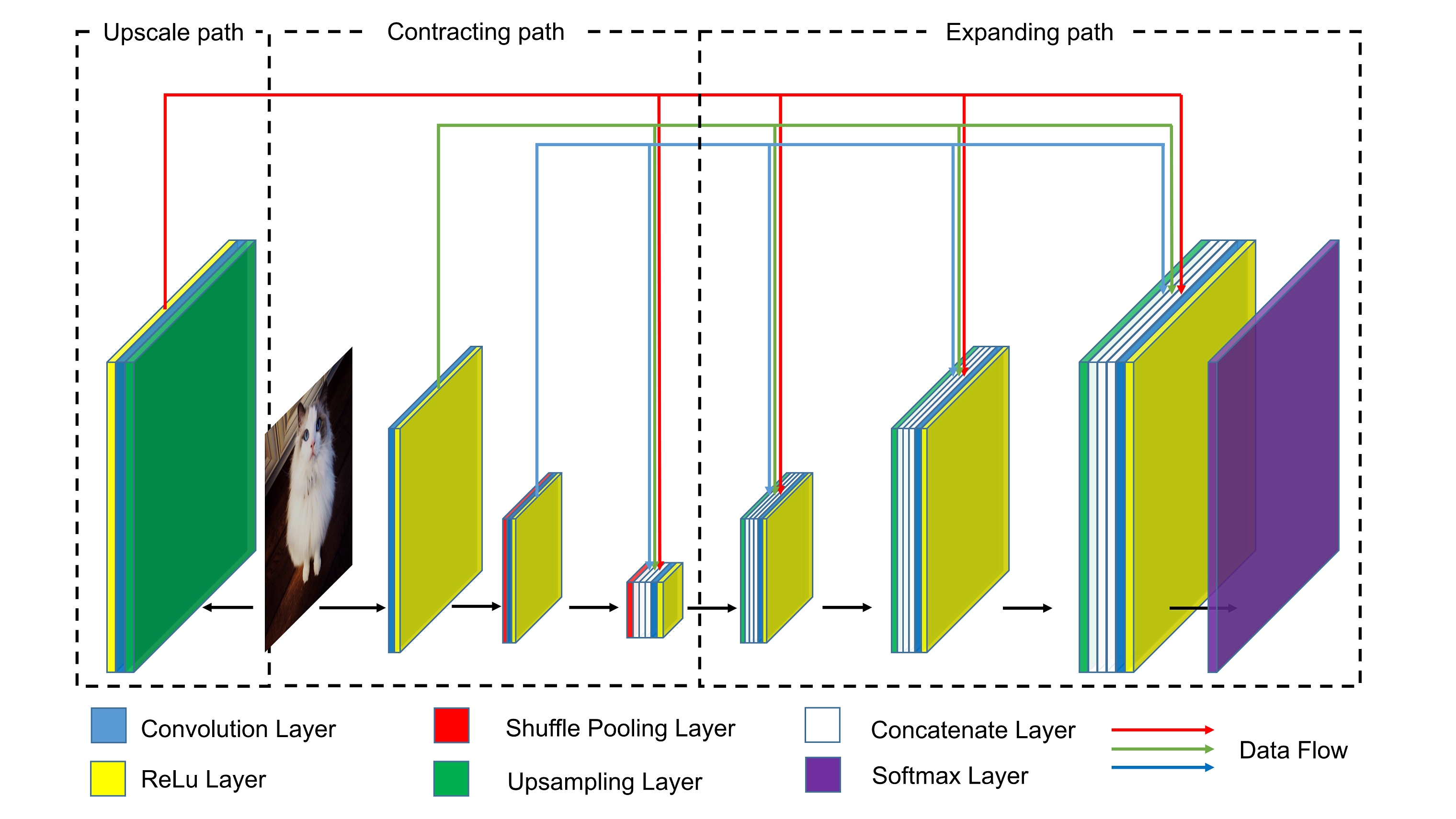}}
\caption{Architecture of Dense U-net for super-resolution task, where the depth of the network is 3 and the up-scale is 2.}
\label{fig:CNNArc}
\end{figure*}

\subsection{Overall Framework}

The overall pipeline of our approach is shown in Fig.~\ref{fig:CNNArc}. 
the network consists of three main parts: upscale path for the pre-upsampling, contracting path for the feature extraction and expanding path for the image reconstruction.
The proposed {\bf Dense U-net} is taken as the backbone of the network. 
The novel shuffle pooling strategy is designed to replace traditional pooling methods.

\subsection{Dense U-net}

The {\bf Dense UnetSR} for super-resolution is an improved {\bf UnetSR}~\cite{lu2019single} by combining dense blocks into the network. As illustrated in Fig.~\ref{fig:CNNArc}, the network consists of four parts:

\subsubsection{Contracting path}
The left side of {\bf Dense UnetSR} is the contracting path for extracting features. The contracting path contains the continues part of one $3\times 3$ kernel, followed by a Rectified Linear Unit (ReLU) layer, and then a $2\times 2$ max-pooling operation with stride 2 for down-sampling. 
The U-Net~\cite{ronneberger2015u} is modified for SISR task by removing all batch normalization layers and one convolution layer in each block. 
To improve the down-sampling method, shuffle pooling method is applied into {\bf Dense UnetSR} to replace the max-pooling method.

\subsubsection{Expanding path}
The right side of Fig.~\ref{fig:CNNArc} is expanding path for decoding. Each block in the expansive includes an up-sampling of the feature map, which followed by a $2\times 2$ kernel that halves the number of feature maps, and one $3\times 3$ convolution kernel, followed by a ReLU layer.

\subsubsection{Upscale path}

The upscale path includes an up-sampling layer and a convolutional layer and meant to keep the same depth of contracting path and expanding path.
The input image is up-sampling by {\bf bicubic} interpolation and builds the symmetric feature extraction layer corresponds to the same depth up-sampling layer in expanding path.

\subsubsection{Dense skip connection}

The dense data skip-connections are constructed for transferring feature maps from all depth of contracting blocks into every expanding block. 
Because the up-sampling block combines all down-sampling feature maps from all depth of contracting blocks instead of feature maps from the one-way down-sampling path, theoretically the dense skip connection establishes a multi-path data transmission and reduces the information transmission loss.

\subsection{Shuffle pooling}

\begin{figure}[H]
\centerline{\includegraphics[width=0.45\columnwidth]{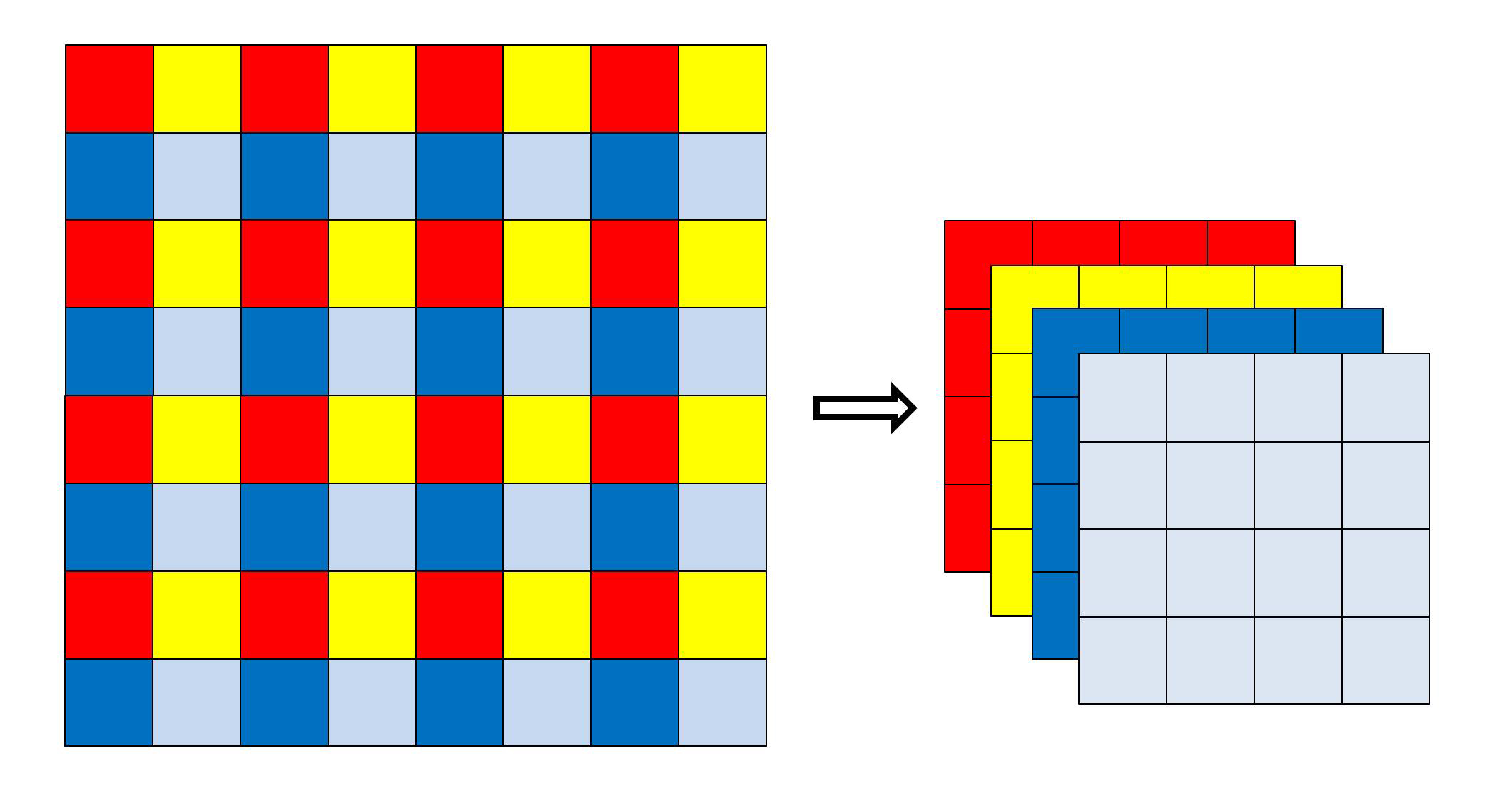}}
\caption{Shuffle pooling flow diagram, where the \\down-scale is 2.}
\label{fig:shufflepool}
\end{figure}

\begin{figure}[H]
    \centering
    \subfloat[Direct method]{\label{fig:reshapepooling}
    \includegraphics[width=0.45\columnwidth]{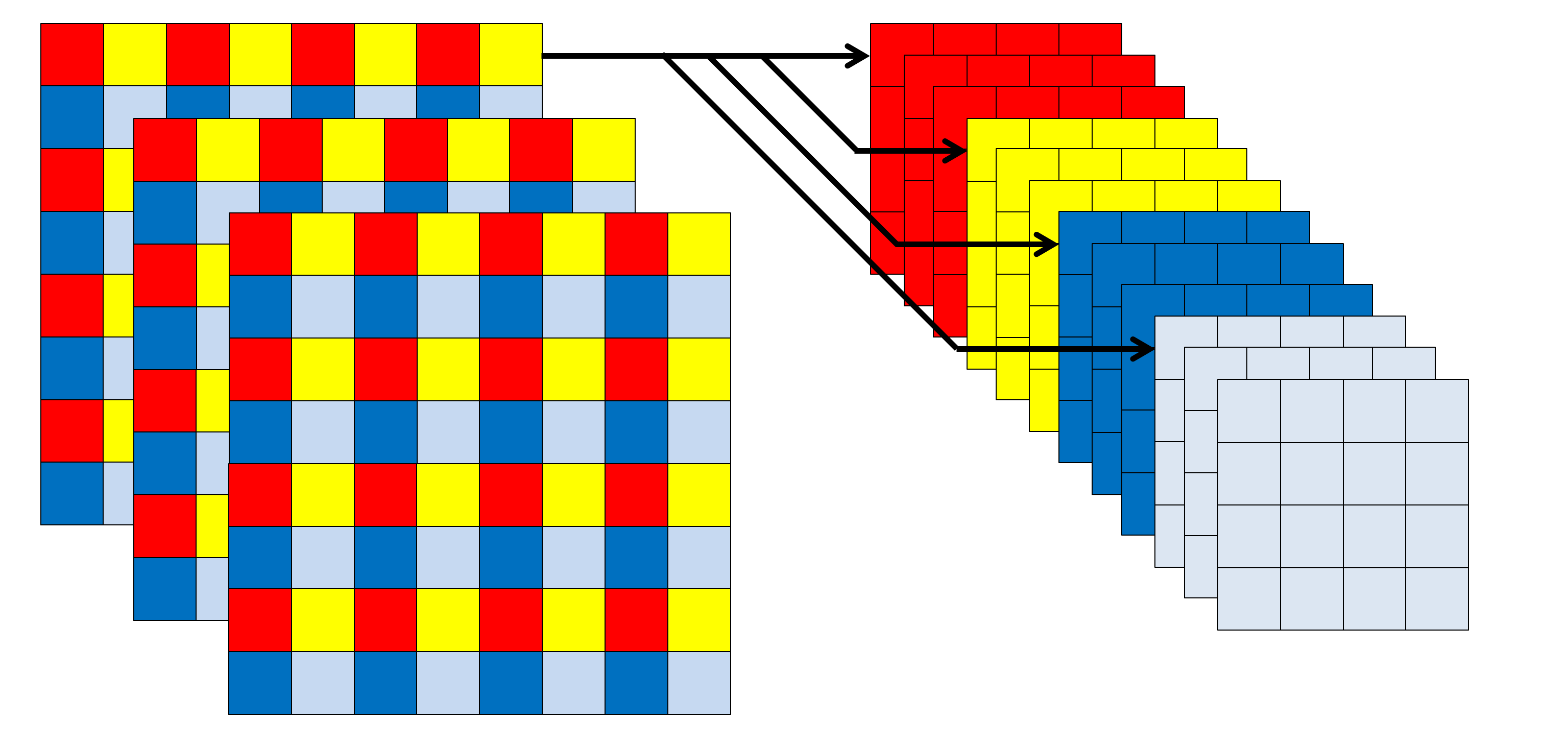}}
    \quad
    \subfloat[Insert method]{\label{fig:shufflepooling}
    \includegraphics[width=0.45\columnwidth]{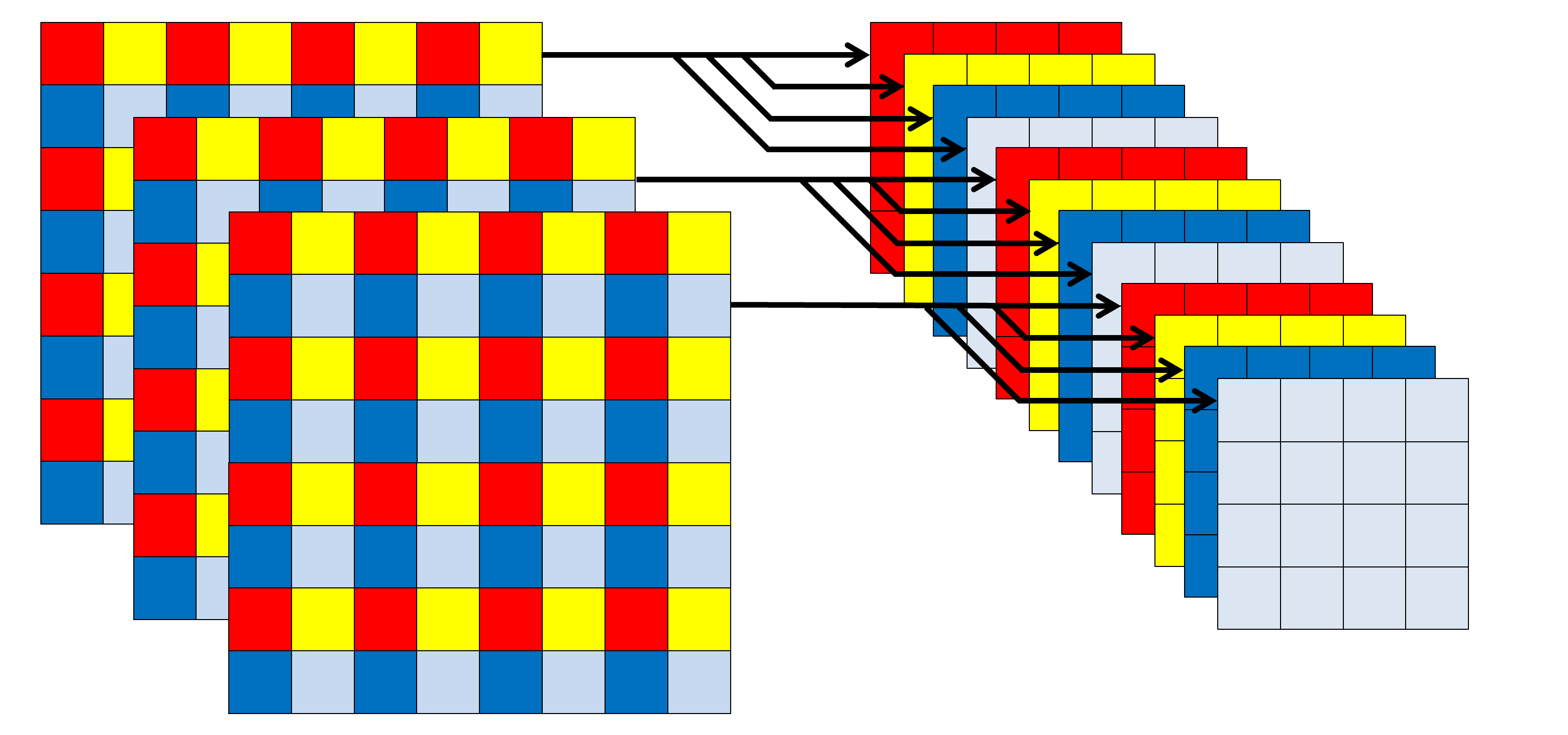}}\\

    \caption{Different arrangements of pooling methods, where the down-scale is 2.}
    \label{fig:diffarrange}
\end{figure}

In most network architectures, general pooling methods including max-pooling and average-pooling were widely applied for down-sampling. 
In this work, a new pooling strategy called shuffle pooling is presented. 
As illustrated in Fig.~\ref{fig:shufflepool}, this module consists of 2 steps: 1) Sampling the input feature map by a specified interval; 2) Reshaping sampled values into down-sampled feature maps.
According to the different arrangement of pooling layer, the shuffle pooling have two different strategies, namely directive and insertive shuffle pooling, which are shown in Fig.~\ref{fig:diffarrange}. Different from the max-pooling and average-pooling which keep one-quarter sampling, the proposed shuffle pooling maintains all information in the feature map of the previous layer. Meanwhile, shuffle pooling brings four times the channel number and four times the parameter amount.

\subsection{Mix loss function}

In this subsection, we first analysize the problem of existing SISR which were only trained with Mean Square Error (MSE) loss. Then the three losses used for the proposed mix loss function, namely MSE loss,  Structural Similarity Index (SSIM) loss and Mean Gradient Error (MGE) loss are introduced respectively. The mix loss function is presented at the end of the subsection.

\subsubsection{Problem analysis}

\begin{figure}[H]
\centerline{\includegraphics[width=0.75\columnwidth]{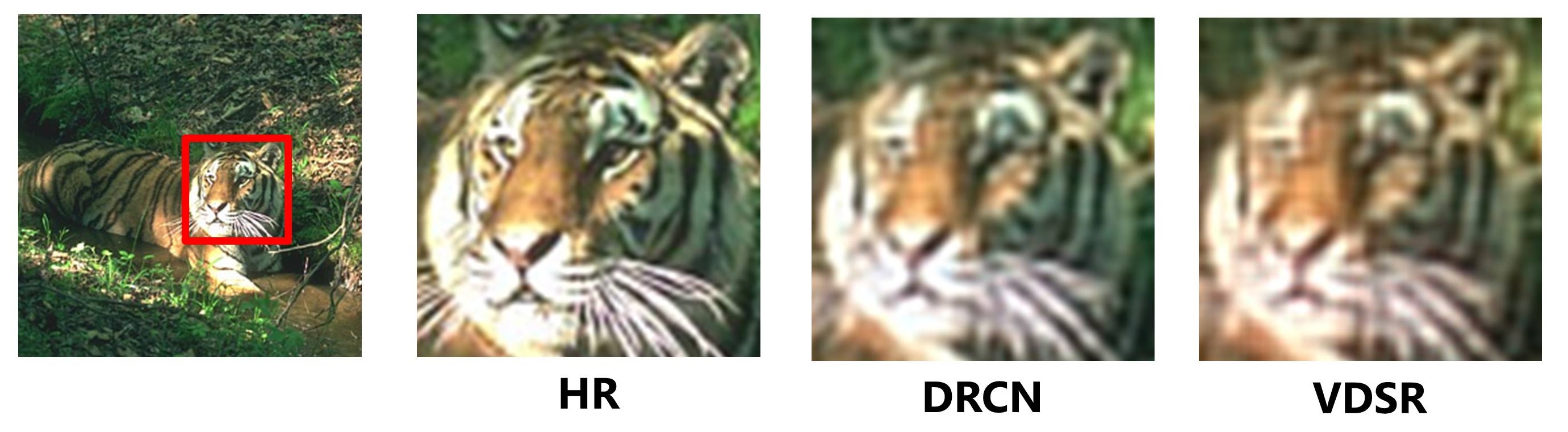}}
\caption{Super-resolution results by previous works with MSE loss($\times$ 4).}
\label{fig:previous}
\end{figure}

As shown in Fig.~\ref{fig:previous}, most previous works on SISR task only trained by MSE loss function, therefore these methods had some shortcomings.

First, previous studies of SISR trained only by MSE loss had not dealt with the problem of blurred edge,
shown as reconstructed results by {\bf DRCN} and  {\bf VDSR} in Fig.~\ref{fig:previous}.
To preserve the sharpness of the reconstructed edges, it is necessary to take image gradient as one of the SISR constraints.

Second, MSE only focused on the error between each pixel and ground truth, which ignores the neighboring structure of pixels.
SSIM gave us a new method to measure the structural similarity between reconstructed images and the ground truth, which can be applied in SISR task as a  neighboring structure constraint to improve the structural similarity.

\subsubsection{MSE loss}

MSE reflects the variance between the current image and the source image at each pixel. 
Two criteria are described as follows. Let \emph{Y} donates the ground truth and \emph{$\hat{Y}$} donates the reconstructed high-resolution images respectively.

\begin{equation}
\begin{split}
MSE=\frac{1}{MN}\sum_{i=1}^{M}\sum_{j=1}^{N} (\hat{Y}(i,j)-Y(i,j))^{2}
\end{split}
\end{equation}

\subsubsection{SSIM loss}

The criterion of SSIM between patches $P_{\hat{Y}}$ and $P_{Y}$ at the same location on ground truth images $\hat{Y}$ and reconstructed high-resolution image $Y$ is defined as

\begin{equation}
\begin{split}
SSIM(P_{\hat{Y}},P_{Y})=\frac{(2\mu_{ P_{\hat{Y}}}\mu_{P_{Y}}+c_{1})(2\sigma_{ P_{\hat{Y}}}\sigma_{  P_{Y}}+c_{2})}{(\mu_{P_{\hat{Y}}}^{2}+\mu_{P_{Y}}^{2}+c_{1})(\sigma_{P_{\hat{Y}}}^{2}+\sigma_{P_{Y}}^{2}+c_{2})}
\end{split}
\end{equation}

\noindent where $\mu_{P_{\hat{Y}}}$ and $\mu_{P_{Y}}$ are the mean of patch $P_{\hat{Y}}$ and $P_{Y}$ respectively. Meanwhile, $\sigma_{P_{\hat{Y}}}$ and $\sigma_{P_{Y}}$ are the deviation of patch $P_{\hat{Y}}$ and $P_{Y}$. $c_{1}$ and $c_{2}$ are small constants. Then, The criterion of $SSIM(\hat{Y} ,Y)$ is the average of patch-based SSIM over the image.

\subsubsection{MGE loss}

To solve the gradient error measurement problem, we introduce classic gradients to the SISR loss function called Sobel operator\cite{kanopoulos1988design}. 
The gradient map $G$ in $x$ and $y$ direction of the ground truth image $Y$ shows below:

\begin{equation}
G_{x}=Y*
\begin{bmatrix}
-1 & -2 & -1\\ 
0 &  0 & 0\\ 
1 &  2 & 1
\end{bmatrix}
\end{equation}

\begin{equation}
G_{y}=Y*
\begin{bmatrix}
-1 &  0 & 1\\ 
-2 &  0 & 2\\ 
-1 &  0 & 1
\end{bmatrix}
\end{equation}

\noindent where $*$ is the convolution operation.

Then we combine the gradient value of $x$ and $y$ direction as follows:

\begin{equation}
G(i,j)=\sqrt{G_{x}^{2}(i,j)+G_{y}^{2}(i,j)}
\end{equation}

Let \emph{G} donates the ground truth and \emph{$\hat{G}$} donates the reconstructed high-resolution results.
The aim of measure the gradient error is to learn a sharp edge which is close to the ground truth edge. The Mean Gradient Error (MGE) shows as follows:

\begin{equation}
\begin{split}
MGE=\frac{1}{n}\frac{1}{m}\sum_{i=1}^{n}\sum_{j=1}^{m}(\hat{G}(i,j)-G(i,j))^{2}
\end{split}
\end{equation}

\subsubsection{Mix loss function}

After achieving the MGE, it should be emphasized in the mix gradient error. 
As the main component of mix gradient error, Mean Square Error forms a Mix Error (MixE) by adding Mean Gradient Error with a weight of $\lambda_{G}$ and SSIM with a weight of $\lambda_{S}$.

\begin{equation}
\begin{split}
MixE(Y,\hat{Y})=MSE+\lambda_{G}MGE+\lambda_{S}SSIM
\end{split}
\end{equation}

Previous researches on SISR has tended to focus on network architecture rather than loss function.
Therefore, the modified form of the loss function is applied to this experiment.
To illustrate the superiority of mix loss function, a much more systematic inference would try to identify how network performance interacts with MixE that is believed to be linked to the fusion method.

\section{Experiment}

\subsection{Dataset}

ICDAR2003~\cite{karatzas2013icdar} is a dataset of the ICDAR Robust Reading Competition for text detection and recognition tasks. 
Though the ICDAR2003 dataset is not commonly used for SISR task, it reflects the performance of these methods on text images. 
The ICDAR2003 dataset consists of 258 training images and 249 testing images, which contains texts in most of the common life complex circumstances. 
Because of the resolution of images varies from 422$\times$102 to 640$\times$480, we resize them into 224$\times$224 with {\bf bicubic} interpolation. 
This network is also compared with other existing methods over standard benchmark datasets: SET14~\cite{bevilacqua2012low}, BSD300~\cite{martin2001database}.

\subsection{Evaluation method}

Two widely used evaluation methods, Peak Signal to Noise Ratio (PSNR)~\cite{huynh2008scope} and Structural Similarity Index (SSIM)~\cite{hore2010image}, are applied for comparison on image quality and similarity. 

PNSR, which derived from MSE, reflects the ratio of peak signal to noise. 
The criteria of PSNR and SSIM are all based on luminance. The higher the value of these criteria, the better the performance of image reconstruction. 
Compared with MSE, the value of PSNR is positively correlated with image quality, which is more conducive to intuitive comparison.

\begin{equation}
\begin{split}
PSNR(Y,\hat{Y})=10log_{10}\frac{255^{2}}{MSE}
\end{split}
\end{equation}

\subsection{Implement details}

Three datasets,  SET14~\cite{bevilacqua2012low}, BSD300~\cite{martin2001database} and ICDAR2003~\cite{karatzas2013icdar}, are chosen to evaluate these existed SISR methods. SET14 and BSD300 dataset consist of natural scenes and ICDAR2003 contain various types of texts in a robust common scene.
In order to generate low-resolution and high-resolution image pairs for training and testing, the source images are down-scaled by {\bf bicubic} interpolation on Table~\ref{tab1}.
Meanwhile, all images are converted into RGB colour space. 

\begin{table}[th]
\caption{Image size of different scales}
\label{table}
\setlength{\tabcolsep}{3pt}
\centering
\begin{tabular}{|l|c|c|l|}
\hline
\multirow{2}{*}{Scale} & \multicolumn{2}{|c|}{Image size} \\
\cline{2-3}
	&LR	&HR\\
\hline
$\times$2	&	112$\times$112 	&224$\times$224\\
$\times$4	&	56$\times$56 	&224$\times$224\\
$\times$8	&	28$\times$28 	&224$\times$224\\
\hline
\end{tabular}
\label{tab1}
\end{table}

The MSE, which widely used in image reconstruction tasks, is chosen to be the loss function in this experiment.
Two widely used evaluation methods, PSNR~\cite{huynh2008scope} and Structural Similarity Index (SSIM)~\cite{hore2010image}, are applied for comparison on image quality and similarity in our experiment.

In the training parameter set, the batch of data is set to $1$. Our method is trained by Adam optimizer~\cite{kingma2014adam} with $\beta_{1}=0.9$, $\beta_{2}=0.999$, $\epsilon=10^{-8}$. The learning rate is set to $10^{-3}$ initially and decreases to half every $50$ epochs. The PyTorch implement of our method is trained with one RTX 2080 GPU and this project is proposed online.


\subsection{Network analysis}

\subsubsection{Dense U-net}
To evaluate U-net and Dense U-net architecture, we train these methods on U-net and Dense U-net backbone with 5-layer depth over SET14, BSD300 and ICDAR2003 datasets. 
Data from the Table.~\ref{tab4pool} shows that the backbone of Dense U-net performs much better than the U-net on all three datasets. 
Because the insert method of shuffle pooling is better than the direct method from the Table.~\ref{tab4pool},
the insert method is applied into all the following shuffle pooling method. 
U-net for super-resolution is simplified as {\bf UnetSR} and Dense U-net for super-resolution without shuffle pooling is {\bf DensetSR}. 
Meanwhile, Dense U-net with shuffle pooling layer is simplified as {\bf DensetSR+}.

\begin{table}
\caption{Comparison of different shuffle pooling arrangements}
\label{table}
\centering
{\begin{tabular}{lccc}

Scale=2\\
\hline
\hline
\multirow{2}{*}{Method} & \multicolumn{1}{c}{SET14}	& \multicolumn{1}{c}{BSD300}	& \multicolumn{1}{c}{ICDAR2003}\\
			&	PSNR/SSIM		&			PSNR/SSIM	&			PSNR/SSIM\\
\hline
UnetSR$^{\mathrm{a}}$				&	26.7241/0.8889		&	29.4241/0.8832	&	35.8147/0.9307	\\
UnetSR(Direct$^{\mathrm{b}}$)			&	27.1581/0.8984	&	29.5013/0.8854	&	35.8964/0.9312	\\
UnetSR(Insert$^{\mathrm{c}}$)  		&	27.3221/0.9014		&	29.7188/0.8902	&	36.4241/0.9392	\\
DenseSR								&	27.4011/0.9027		&	29.4902/0.8851	&	35.9829/0.9325	\\
DenseSR(Direct)						&	27.4278/0.9033	&	29.5133/0.8858	&	35.9844/0.9327	\\
DenseSR(Insert)						&	28.3938/0.9236		&	29.8478/0.8965	&	36.9244/0.9409	\\
\hline
\hline
\multicolumn{4}{p{230pt}}{$^{\mathrm{a}}$The max-pooling method is default to UnetSR and DenseSR.}\\
\multicolumn{4}{p{230pt}}{$^{\mathrm{b}}$This is the direct method of shuffle pooling.}\\
\multicolumn{4}{p{230pt}}{$^{\mathrm{c}}$This is the insert method of shuffle pooling.}\\

\end{tabular}}
\label{tab4pool}
\end{table}

\subsubsection{Shuffle pooling}
As Table.~\ref{tab4pool} shown, there is a significant improvement by replacing the max-pooling method with shuffle pooling. A comparison of the two arrangement methods in Table.~\ref{tab4pool} reveals the insert method of shuffle pooling achieves the higher PSNR than the direct method, with +0.94 dB on SET14 dataset, +0.97 dB on BSD300 dataset and  +1.10 dB on ICDAR2003 dataset.

\subsubsection{Mix loss function}

\begin{figure*}[h]
\centerline{\includegraphics[width=0.5\columnwidth]{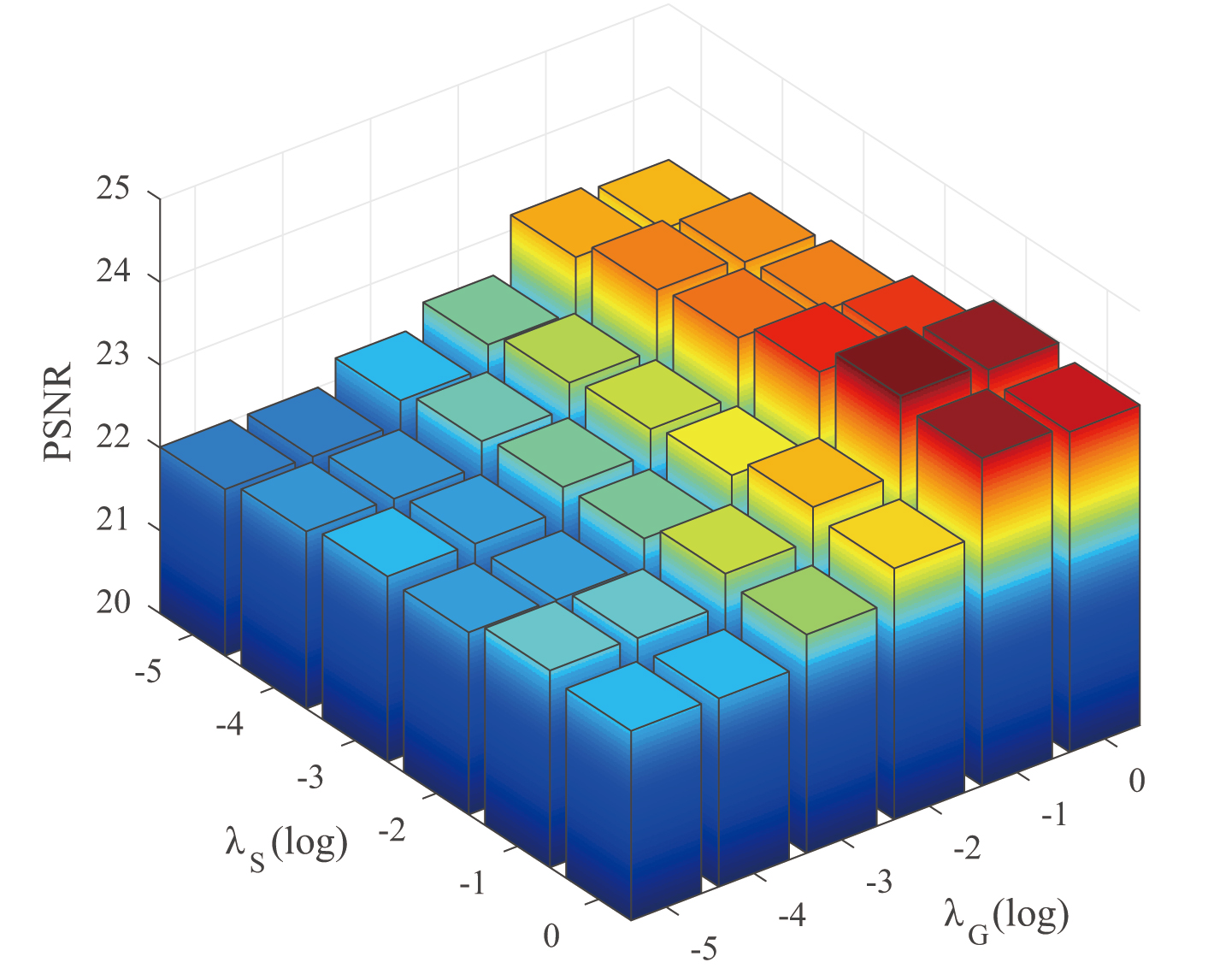}}
\caption{An ablation experiment of $\lambda_{G}$ and $\lambda_{S}$ over BSD300 dataset($\times 8$).}
\label{fig:Loss4lambda}
\end{figure*}

In order to determine the best value of the weight, the result of an ablation experiment on the scale of 8 over BSD300 dataset is proposed in Fig.~\ref{fig:Loss4lambda}.

The main point in Fig.~\ref{fig:Loss4lambda} to note is that MGE has a greater effect on PSNR than SSIM. 
A reliable explanation is that they have serious homogeneity because SSIM and MSE are both in the RGB domain, but MGE is in the gradient domain and provides a cross-domain information.
Under the guidance of the ablation experiment result, $\lambda_{G}$ is set to $0.1$ and $\lambda_{S}$ is set to $0.1$ in all subsequent experiments.

\begin{table}[H]
\caption{Evaluation of Mix Loss on U-net and Dense U-net backbone over SET14, BSD300 and ICDAR2003 datasets}
\label{table}
\centering
{\begin{tabular}{llccccccl}
Scale=2\\
\hline
\hline
\multirow{2}{*}{Method} & \multirow{2}{*}{Loss} &\multicolumn{1}{c}{SET14}&\multicolumn{1}{c}{BSD300} &\multicolumn{1}{c}{ICDAR2003}			\\
		&	&	PSNR/SSIM		& 	PSNR/SSIM	& 	PSNR/SSIM		\\
\hline
UnetSR 		&	MSE		&	26.7241/0.8889	&	29.4241/0.8832	&	35.8147/0.9307	\\
UnetSR 		&	MixE	&	27.1359/0.8974	&	29.6336/0.8898	&	36.0978/0.9364	\\
DenseSR		&	MSE		&	27.4011/0.9027	&	29.4902/0.8851	&	35.9829/0.9325	\\
DenseSR		&	MixE	&	27.7713/0.9106	&	29.9828/0.9009	&	36.2948/0.9382	\\
\hline
\hline
\\
Scale=4\\
\hline
\hline
\multirow{2}{*}{Method} & \multirow{2}{*}{Loss}& \multicolumn{1}{c}{SET14}&\multicolumn{1}{c}{BSD300} &\multicolumn{1}{c}{ICDAR2003}			\\
			&	&	PSNR/SSIM		& 	PSNR/SSIM		& 	PSNR/SSIM		\\
\hline
UnetSR		&	MSE		&	20.8891/0.6693	&	24.8332/0.6843	&	29.3374/0.8202	\\
UnetSR		&	MixE		&	21.3520/0.6859	&	25.1819/0.7033	&	29.8233/0.8181	\\
DenseSR 	&	MSE		&	21.4532/0.7009	&	24.9723/0.6927	&	30.4351/0.8392	\\
DenseSR		&	MixE		&	21.9932/0.7181	&	25.4017/0.7141	&	31.0083/0.8523	\\
\hline
\hline
\\
Scale=8\\
\hline
\hline
\multirow{2}{*}{Method} & \multirow{2}{*}{Loss}& \multicolumn{1}{c}{SET14}&\multicolumn{1}{c}{BSD300} &\multicolumn{1}{c}{ICDAR2003}			\\
			&&	PSNR/SSIM		& 	PSNR/SSIM		& 	PSNR/SSIM		\\
\hline
UnetSR 		&	MSE		&	16.7001/0.4093	&	21.9865/0.5231	&	25.7734/0.7105	\\
UnetSR 		&	MixE	&	16.9707/0.4177	&	22.2273/0.5254	&	25.8770/0.7148	\\
DenseSR 	&	MSE		&	18.9354/0.5752	&   	24.0628/0.6591	&   	29.0465/0.8051	\\
DenseSR 	&	MixE	&	19.5080/0.6158	&	24.4807/0.6637	&	29.5169/0.8357	\\
\hline
\hline
\multicolumn{4}{p{300pt}}{The MSE and MixE are only applied for network training.}\\
\multicolumn{4}{p{300pt}}{The MSE and MixE are independent of evaluation methods.}\\
\end{tabular}}
\label{tab4loss}
\end{table}

The evaluation of mix loss function on the backbone of Dense U-net in Table.~\ref{tab4loss} illustrates its effectiveness.
The modification of loss function obtains a great improvement on PSNR in most instances, which manifests in approximately +0.8dB increment.

\subsection{Comparison with the state-of-the-arts}

\begin{figure}[h]
    \centering
    \subfloat[Super-resolution results($4\times$) of 182053.jpg from BSD300 dataset]{\label{fig:reshapepooling}
    \includegraphics[width=1\columnwidth]{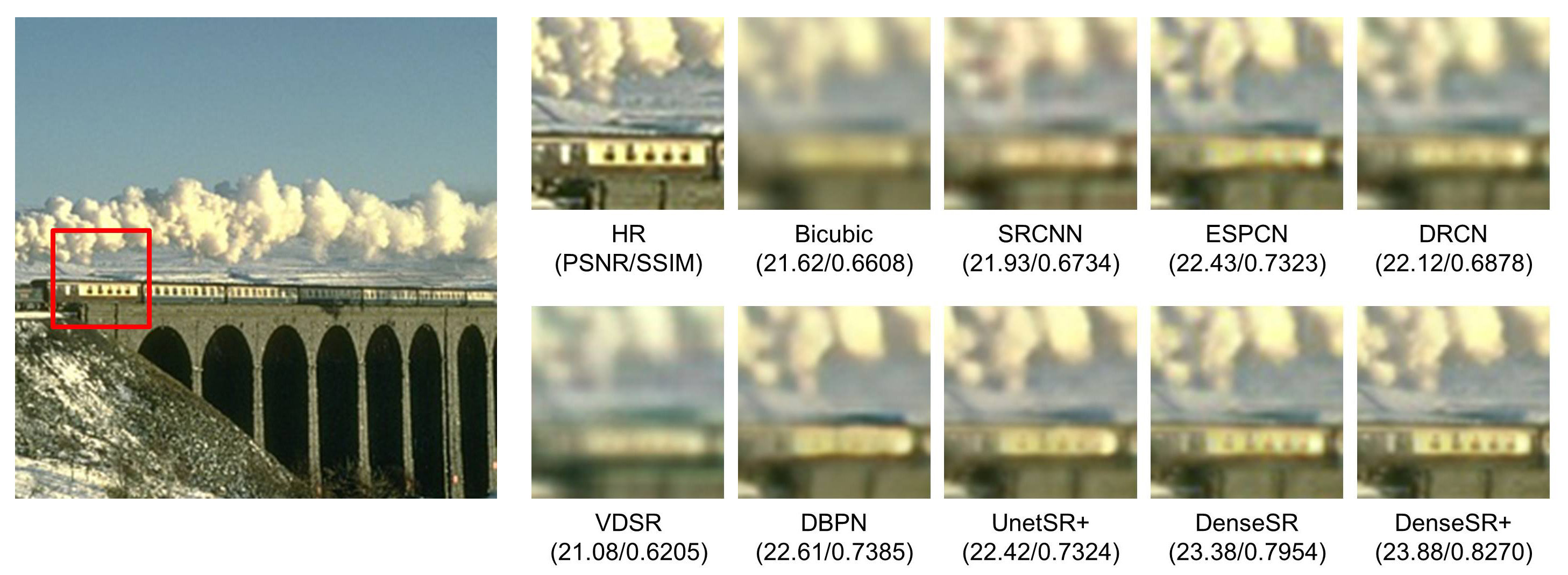}}\\
    \subfloat[Super-resolution results($2\times$) of 148026.jpg from BSD300 dataset]{\label{fig:shufflepooling}
    \includegraphics[width=1\columnwidth]{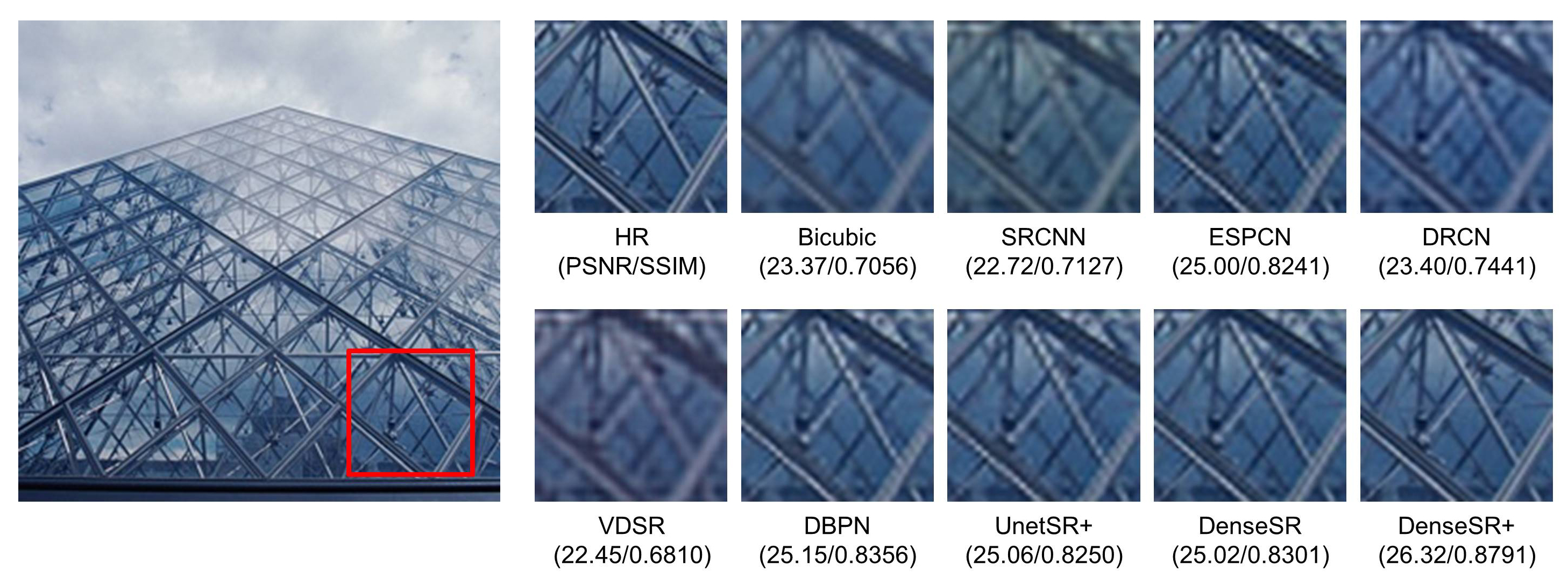}}\\

    \caption{Reconstructed images by different methods.}
    \label{fig:demo}
\end{figure}

The comparison between our method and previous works, includes {\bf ESPCN}\cite{shi2016real}, {\bf SRCNN}\cite{dong2015image}, {\bf VDSR}\cite{kim2016accurate}, {\bf EDSR}\cite{lim2017enhanced}, {\bf FSRCNN}\cite{dong2016accelerating}, {\bf DRCN}\cite{kim2016deeply}, {\bf SRGAN}\cite{ledig2017photo}, {\bf DBPN}\cite{haris2018deep}, {\bf RCAN}\cite{zhang2018image}, {\bf SAN}\cite{dai2019second}  and {\bf Bicubic}\cite{de1962bicubic}, are shown as Table.~\ref{tab3} and Fig.~\ref{fig:demo}(a)(b). In Table.~\ref{tab3}, red numbers mark the highest score and blue numbers mark the second best results. It can be clear seen that the {\bf DenseSR+} performs the best under most circumstances. 
Table.~\ref{tab4complex} shows the comparison of parameter numbers and running time between different deep learning methods.
Though {\bf DenseSR+} is a large parameter number deep learning method, it achieves the state-of-the-art performance.

\begin{table}[!h]

\caption{Comparison of PSNR and SSIM with the parameter number and running time}
\label{table}
\centering
{\begin{tabular}{lrrcccccl}

Scale=8\\
\hline
\hline
\multirow{2}{*}{Method} & \multirow{2}{*}{Params}&\multirow{1}{*}{Time}&\multicolumn{1}{c}{BSD300} &\multicolumn{1}{c}{ICDAR2003}			\\
						&		&(ms)	& 	PSNR/SSIM		& 	PSNR/SSIM		\\
\hline
Bicubic\cite{de1962bicubic}	&			&							&	21.3115/0.4933	&	24.3856/0.6831	\\
ESPCN\cite{shi2016real} 		&	75552	&	1.2980 			&	21.6447/0.5064	&	25.1132/0.6964	\\
SRCNN\cite{dong2015image}	&	171488	&	3.0600 			&	21.8101/0.5075	&	22.6281/0.6103	\\
VDSR\cite{kim2016accurate} 	&	224640	&	6.9120 			&	21.9697/0.5181	&	25.6303/0.7104	\\
EDSR\cite{lim2017enhanced} 	&	779523	&	16.540 			&	21.6573/0.5067	&	23.5578/0.5987	\\
FSRCNN\cite{dong2016accelerating}	&	27267	&	0.8890 	&	21.3311/0.5011	&	22.5721/0.6155	\\
DRCN\cite{kim2016deeply}	&	114307	&	60.5730 				&	21.2771/0.4934	&	24.2561/0.6725	\\
SRGAN\cite{ledig2017photo} 	&	6535494	&	12.0300 			&	21.8766/0.5121	&	23.5621/0.6425	\\
DBPN\cite{haris2018deep} 	&	23205878	&	82.5170 			&	22.0577/0.5229	&	26.3482/0.7196	\\
RCAN\cite{zhang2018image}	&	37129404	&	43.2140		&	22.5201/0.5712	&	{\color{blue}29.3084}/0.7928	\\
SAN\cite{dai2019second} 		&	30248124	&	106.3800	&	{\color{blue}24.1560}/{\color{blue}0.6701}	&	28.6021/0.7344	\\
UnetSR~\cite{lu2019single}	&	8495907	&	16.9530 			&	21.9865/0.5231	&	25.7734/0.7105	\\
DenseSR					&	19109451&	18.2470  			&	24.0628/0.6591	&	29.0465/{\color{blue}0.8051}	\\
DenseSR+					&	121352587&	126.7016  		&	{\color{red}25.5143}/{\color{red}0.6874}	&{\color{red}30.6092}/{\color{red}0.8413}	\\

\hline
\hline
\multicolumn{4}{p{200pt}}{Red numbers mark the best score.}\\
\multicolumn{4}{p{200pt}}{Blue numbers mark the second best score.}\\
\end{tabular}}
\label{tab4complex}
\end{table}

\begin{table}[htbp]
\caption{Comparison results on scale of 4}
\label{table}
\setlength{\tabcolsep}{3pt}
\centering
{\begin{tabular}{lcccccccccl}

Scale=2\\
\hline
\hline
\multirow{2}{*}{Method} & \multicolumn{1}{c}{SET14}&\multicolumn{1}{c}{BSD300} &\multicolumn{1}{c}{ICDAR2003}			\\
			&	PSNR/SSIM		& 	PSNR/SSIM		& 	PSNR/SSIM		\\
\hline

Bicubic\cite{de1962bicubic}	&	24.4523/0.8482	&	26.6538/0.7924	&	32.9327/0.9028	\\
ESPCN\cite{shi2016real} 		&	26.7606/0.8999	&	28.9832/0.8732	&	35.6041/0.9243	\\
SRCNN\cite{dong2015image}	&	25.9711/0.8681	&	28.6943/0.8671	&	35.2711/0.9234	\\
VDSR\cite{kim2016accurate} 	&	28.6617/0.9269	&	29.3889/0.8785	&	36.2323/0.9375	\\
EDSR\cite{lim2017enhanced} 	&	24.0624/0.8383	&	28.3119/0.8621	&	34.5047/0.9258	\\
FSRCNN\cite{dong2016accelerating}&	23.1284/0.8123	&	28.7534/0.8681	&	35.0533/0.9355	\\
DRCN\cite{kim2016deeply}		&	24.4234/0.8458	&	27.5089/0.8088	&	33.7849/0.9185	\\
SRGAN\cite{ledig2017photo} 	&	23.9553/0.8195	&	28.7072/0.8633	&	33.2834/0.9135	\\
DBPN\cite{haris2018deep} 		&	28.4092/0.9202	&	29.8675/0.8834	&	36.2344/{\color{blue}0.9401}	\\
RCAN\cite{zhang2018image}	&	28.9150/0.9271	&	29.7674/0.8878	&	36.8491/0.9377	\\
SAN\cite{dai2019second} 		&	{\color{blue}28.9672}/{\color{blue}0.9298}	&	{\color{blue}29.9189}/{\color{blue}0.9005}	&	{\color{blue}36.8947}/0.9387	\\
UnetSR~\cite{lu2019single}	&	26.7241/0.8889	&	29.4241/0.8832	&	35.8147/0.9307	\\
DenseSR						&	27.4011/0.9027	&	29.4902/0.8851	&	35.9829/0.9325	\\
DenseSR+					&{\color{red} 29.1197}/{\color{red}0.9331}	&{\color{red} 30.2264}/{\color{red}0.9086}	&{\color{red}37.4785}/{\color{red}0.9487}	\\
\hline
\hline
Scale=4\\
\hline
\hline
\multirow{2}{*}{Method} & \multicolumn{1}{c}{SET14}&\multicolumn{1}{c}{BSD300} &\multicolumn{1}{c}{ICDAR2003}			\\
			&	PSNR/SSIM		& 	PSNR/SSIM		& 	PSNR/SSIM		\\
\hline

Bicubic\cite{de1962bicubic}	&	19.7167/0.6089	&	23.5053/0.6157	&	28.1135/0.7875	\\
ESPCN\cite{shi2016real} 		&	20.6292/0.6333	&	24.4899/0.6641	&	29.4861/0.8214	\\
SRCNN\cite{dong2015image}		&	20.5825/0.6288	&	24.2232/0.6597	&	28.1906/0.7661	\\
VDSR\cite{kim2016accurate} 	&	21.4763/0.6991	&	24.7077/0.6816	&	30.5267/0.8321	\\
EDSR\cite{lim2017enhanced} 	&	19.9784/0.6269	&	23.9192/0.6513	&	27.9723/0.7101	\\
FSRCNN\cite{dong2016accelerating}&	19.3255/0.5941	&	24.2499/0.6599	&	28.0231/0.7652	\\
DRCN\cite{kim2016deeply}		&	19.7077/0.6078	&	23.3462/0.6132	&	27.7174/0.7764	\\
SRGAN\cite{ledig2017photo} 	&	19.3877/0.5976	&	24.1675/0.6485	&	27.5605/0.7654	\\
DBPN\cite{haris2018deep} 		& 	21.7657/0.7171	&	25.0644/0.6967	&	29.8832/0.8224	\\
RCAN\cite{zhang2018image}	&	21.7085/0.7160	&	25.4241/0.6976	&	{\color{blue}30.8147}/{\color{blue}0.8407}	\\
SAN\cite{dai2019second} 		&	{\color{blue}21.9241}/{\color{blue}0.7240}	&	{\color{red}25.9141}/{\color{red}0.7401}	&	29.7817/0.8121	\\
UnetSR~\cite{lu2019single}		&	20.8891/0.6693	&	24.8332/0.6843	&	29.3374/0.8202	\\
DenseSR					&	21.4532/0.7009	&	24.9723/0.6927	&	30.4351/0.8392	\\
DenseSR+					&{\color{red}22.2568}/{\color{red}0.7292}&{\color{blue}25.9024}/{\color{blue}0.7386}	&{\color{red}32.0652}/{\color{red}0.8829}	\\
\hline
\hline	
Scale=8\\
\hline
\hline
\multirow{2}{*}{Method} & \multicolumn{1}{c}{SET14}&\multicolumn{1}{c}{BSD300} &\multicolumn{1}{c}{ICDAR2003}			\\
			&	PSNR/SSIM		& 	PSNR/SSIM		& 	PSNR/SSIM		\\
\hline

Bicubic\cite{de1962bicubic}	&	16.1132/0.3673	&	21.3115/0.4933	&	24.3856/0.6831	\\
ESPCN\cite{shi2016real} 		&	16.3441/0.3628	&	21.6447/0.5064	&	25.1132/0.6964	\\
SRCNN\cite{dong2015image}	&	16.3853/0.3614	&	21.8101/0.5075	&	22.6281/0.6103	\\
VDSR\cite{kim2016accurate} 	&	16.7994/0.4095	&	21.9697/0.5181	&	25.6303/0.7104	\\
EDSR\cite{lim2017enhanced} 	&	15.7257/0.3209	&	21.6573/0.5067	&	23.5578/0.5987	\\
FSRCNN\cite{dong2016accelerating}&	14.5788/0.2541	&	21.3311/0.5011	&	22.5721/0.6155	\\
DRCN\cite{kim2016deeply}		&	16.1497/0.3685	&	21.2771/0.4934	&	24.2561/0.6725	\\
SRGAN\cite{ledig2017photo} 	&	15.7133/0.3221	&	21.8766/0.5121	&	23.5621/0.6425	\\
DBPN\cite{haris2018deep} 		&	16.7398/0.4122	&	22.0577/0.5229	&	26.3482/0.7196	\\
RCAN\cite{zhang2018image}	&	18.7580/0.5425	&	22.5201/0.5712	&	{\color{blue}29.3084}/0.7928	\\
SAN\cite{dai2019second} 		&	{\color{blue}19.0185}/{\color{blue}0.6315}	&	{\color{blue}24.1560}/{\color{blue}0.6701}	&	28.6021/0.7344	\\
UnetSR~\cite{lu2019single}	&	16.7001/0.4093	&	21.9865/0.5231	&	25.7734/0.7105	\\
DenseSR						&	18.9354/0.5752	&24.0628/0.6591		&	29.0465/{\color{blue}0.8051}	\\
DenseSR+					&{\color{red}20.9134}/{\color{red}0.6746}	&{\color{red}25.5143}/{\color{red}0.6874}	&{\color{red}30.6092}/{\color{red}0.8213}	\\

\hline
\hline
\end{tabular}}
\label{tab3}
\end{table}

\section{Conclusion}

To solve the problem of limitations caused by defects in the information transmission structure,
we propose a Dense U-net with shuffle pooling layer for super-resolution tasks and it achieves the state-of-the-art result In this work.
Compare with U-net~\cite{ronneberger2015u} for SISR, Dense U-net reduces the information transmission loss because the up-sampling block combines all down-sampling feature maps from all depth of contracting blocks.
Then, a novel pooling method called shuffle pooling is designed for the Dense U-net, which can effectively replace the handcrafted filter in the SISR pipeline with more lossy down-sampling filters specifically trained for each feature map, whilst also reducing the information loss of the overall SISR operation.
Furthermore, the mix loss function, which combined with Mean Square Error(MSE)~\cite{huynh2008scope}, Structural Similarity Index (SSIM)~\cite{hore2010image} and Mean Gradient Error(MGE)~\cite{lu2019single}, basically solves the perception loss and high-frequency information loss. 
In experiments, the proposed {\bf Dense UnetSR} outperforms the state-of-the-art the SISR methods in SET14, BSD300, ICDAR2003 datasets, especially in the text tasks.

\begin{acknowledgements}
This work is supported by the National Natural Science Foundation of China(grant no. 61573168).
\end{acknowledgements}

\bibliography{report} 
\bibliographystyle{spmpsci}      

\end{document}